\documentclass[structabstract]{aa}  
\usepackage{graphicx}
\usepackage{txfonts}
\begin{document}
    \title{Detecting Planets around Very Cool Dwarfs at Near Infrared Wavelengths
with the Radial Velocity Technique}

\author{F. Rodler\inst{1,2}
\and
C. del Burgo\inst{3} 
\and
S. Witte\inst{4}
\and
Ch. Helling\inst{5}
\and
P. H. Hauschildt\inst{4}
\and
E. L. Mart\'in\inst{6}
\and
C. \'Alvarez\inst{1}}
\offprints{frodler@iac.es}
 \institute{Instituto de Astrof\'{i}sica de Canarias,
             C/V\'{i}a L\'{a}ctea s/n, E-38205 La Laguna, Spain
     \and
Institut de Ciencies de l'Espai, Campus UAB, Torre C5 - parell - 2$^a$ planta, 08193 Bellaterra, Spain
    \and
	     		UNINOVA-CA3, Campus da Caparica, 2825-149 Caparica, Portugal
		\and
		Hamburger Sternwarte, Gojenbergsweg 112, D-21029 Hamburg, Germany
		\and
		SUPA, School of Physics and Astronomy, University of St.Andrews, North Haugh, St.Andrews KY16 9SS, UK
		\and
		Centro de Astrobiolog\'ia (CAB-CSIC), Ctra. Ajalvir km. 4, E-28850
		Torrej\'on de Ardoz, Madrid, Spain
}

  \date{Received ?; accepted ?}
 
 
  \abstract
   {Radial velocity monitoring of very cool dwarfs such as late M- and hot L-dwarfs has become a promising tool to search for rocky planets as well as to follow-up planetary candidates around dwarfs found by transit surveys. 
   These stars are faint at optical wavelengths, as their spectral flux distribution peaks at near-infrared (NIR) wavelengths. For this reason, it is desirable to measure the radial velocities in this wavelength regime. 
   However, in the NIR there are only very few medium- and high-resolution spectrographs available which are mounted at large telescopes. 
   In the near future, high-resolution spectrographs for the NIR will be built, which will allow us to search for rocky planets around cool M-dwarfs and L-dwarfs from radial velocities monitoring.}
   {We investigate the precision that can be attained in radial velocity measurements of very cool dwarfs in the NIR. 
   The goal is to determine in which 
atmospheric window of the Earth's atmosphere the 
highest radial velocity precision can be achieved to help in designing the
next generation of NIR high-resolution
spectrographs. }
   {Stellar atmosphere synthetic models for an M- and an L-dwarf with
temperatures of 2200 K and 1800 K, respectively, and a theoretical spectrum
of the Earth's transmission in the spectral range from 0.9 to 2.5~$\mu$m are used.
We simulate a series of Doppler-shifted spectra observed with different resolving
powers and signal-to-noise ratios, and for different rotational broadenings of the dwarf. For different combinations of the input parameters we recover the radial velocity by means of cross-correlation with a high signal-to-noise ratio template and determine the associate uncertainties.}
   {The highest precision in radial velocity 
   measurements for the cool M-dwarf is found in the $Y$~band around 1.0~$\mu$m, while for the L-dwarf it is determined in the $J$~band around 1.25~$\mu$m.
   We note that synthetic models may lack of some faint absorption features or underestimate their abundances. Conversely, some instrumental/calibration aspects that are not taken into account in our estimations would rise the uncertainties.}
   {}

 
   \maketitle
%

\section{Introduction}
The search for extrasolar planets  has
led to about 550 discoveries\footnote{The Extrasolar Planets Encyclopedia;
http://www.exoplanet.eu}, mostly detected by means of the radial velocity (RV) technique by using high-resolution spectrographs ($R=\lambda / \Delta \lambda \ge 40,000$)
at optical wavelengths. Most discoveries are giant gaseous planets (typically hot Jupiters) of short periods (of a few days) around stars of spectral 
types F and G. In order to 
find rocky and low-mass planets, or two follow up candidates found by means of the transit technique,
two options arise: either increase the precision of current spectrographs at
optical wavelengths to less than $0.1~{\rm m~s^{-1}}$ and to search for this kind of planets
around F-G type stars, or to monitor low-mass stars, such as M- and L-dwarfs, for the
following reasons.

Very cool stars such as M-dwarfs are the most abundant type ($\sim70 \%$) of stars in the solar neighborhood and the Milky Way in
general (Henry at al.~1997).
The effective temperatures and masses of M-dwarfs, respectively, range from 3700-2200~K and 0.5 to 0.07 solar masses from the M0 to M9.5 spectral types. They exhibit prominent absorption features corresponding to strong neutral atoms, H$_2$0, FeH, VO, CO and TiO.
Due to the low masses of these
objects, the reflex motion of the host star due to the gravitational pull of the
extrasolar planet is higher and more easily detectable than for more massive
host stars. Since M-dwarfs are very cool stars in comparison with solar-type stars, short period planets would more
likely be situated in the habitable zone.

M-dwarfs emit most of their energy around $1.1-1.3~\mu {\rm m}$, in the near-infrared
(NIR), while they appear very faint at optical wavelengths. Recently,
Reiners et al.~(2010) investigated the possibilities of searching low-mass planets
around M-dwarfs with effective temperatures 3500K down to 2600K. These authors used synthetic spectra from the PHOENIX model atmosphere code (Hauschildt et al~1999, Allard et al.~2001) for $T_{\rm eff}=3500$~K, 2800~K and 2600~K with $\log g = 4.75$. As result, they found that the best
wavelength range to observe M-dwarfs is in the $Y$ band (i.e. around $1.0~\mu {\rm m}$). 

First attempts of measuring radial velocity variations among very cool M-dwarfs at NIR wavelengths were done by Mart\'in et al.~(2006). Using NIRSPEC
(McLean et al.~1998), they achieved a RV precision of around 300 m~s$^{-1}$ for the M9.5-dwarf LP944-20. Very recently, several research groups have reported high-precision RV measurements taken in the NIR with CRIRES
(K\"aufl et al.~2004), mounted at the UT1/VLT in the Paranal
Observatory of ESO in Chile. Bean et al.~(2010) conducted high-resolution data of 
31 M-dwarfs (spectral types M4-M9), and report an
RV precision of better than $5~{\rm m~s^{-1}}$. Figueira et al.~(2010) took observations of the planetary
candidate TW~Hya and achieved a RV precision better than $10~{\rm m~s^{-1}}$ by adopting telluric lines as
a stable reference. However, the exploration of the RV in the IR is in its infancy, and a revolution is expected in the forthcoming year with the arrival of the new generation
of high-resolution near-infrared spectrographs.

The next cooler class of dwarfs are the L-dwarfs (Kirkpatrick et al. 1999; Mart\'in et al. 1999), 
which have effective temperatures between $\sim2100$~K and 1500~K, and their spectral energy distribution peak around 1.3~$\mu$m.
Their masses are in the range of $\sim$0.075 and 0.02~M$_\odot$. 
 
 Charbonneau et al.~(2007) presented preliminary first results of a radial velocity survey for exoplanets orbiting L-dwarfs 
 using the PHOENIX spectrograph (Hinkle et al. 2003), which is mounted at the Gemini South telescope. 
 They attained RV precision sufficient to detect 
 Jupiter-mass companions in short-period orbits around L-dwarfs. Using NISPEC at Keck II with a spectra resolving power of $R\sim25,000$, Blake et al.~(2010) investigated 59 late M- and L-dwarfs spanning spectral types M8/L0
 to L6 over a period of six years. These authors made use of the prominent CO-forest around around 2.3~$\mu$m in the $K$-band to measure RVs. Although
 they did not find any planetary companions, they demonstrated that their achieved RV precision of about 50~m~s$^{-1}$ for the brighter M-dwarfs
  and about 200~m~s$^{-1}$ for the fainter L-dwarfs is sufficient to detect planets.

In this paper, we investigate the precision that can be attained in RV measurements of a late M-dwarf and an L-dwarf, with an effective temperature of 2200K and 1800K, respectively.
Our aim is to define a frame for the instrumental settings of planned spectrographs of high-resolution in the NIR, which will be devoted to the search for rocky planets around very low-mass stars, and hot brown dwarfs. We focus our study on the four NIR filter bands that lie on high transmission windows of the Earth's atmosphere. In particular, the $Y$, $J$, $H$, and $K$~bands, that are located at wavelengths of $1.0-1.1\mu {\rm m}$, $1.17-1.33~\mu {\rm m}$, $1.5-1.75 ~\mu {\rm m}$ and $2.07-2.35 ~\mu {\rm
m}$, respectively.

This paper is organized as followed: Section 2 is devoted to a brief discussion of the limiting factors to radial velocity measurements. Section 3 provides a description of the simulations and the data analysis, which is followed by Section 4, in which the results are presented.
In Section 5, our results are discussed.

\section{Limitation to radial velocity measurements of late M-dwarfs in the NIR}

In general, the expected RV precision $\sigma_{\rm RV}$ can be estimated by the following equation given in Butler et al.~(1996): 
\begin{equation}       \label{E:-1}
\sigma_{\rm RV}^{-2}= \sum_{i} (\frac{{\rm d}I_i}{{\rm d}V} ~ {\rm SNR}_i)^2 ~,
\end{equation}
where ${\rm d}I_i / {\rm d}V$ represent the differential increment in the spectrum (i.e. the first derivative at pixel $i$), and SNR$_i$ the signal-to-noise ratio at pixel $i$.  Equation~\ref{E:-1} means that for  
a given SNR, the RV precision is higher if the star has a large number of deep and narrow absorption lines. Any 
broadening coming from the star or instrument will cause that the absorption lines in the stellar spectrum get shallower and weaker, 
which leads to a lower precision in the RV measurements. Such a broadening is predefined by the spectral resolving power $R=\lambda/\Delta\lambda$ given by  
the spectrograph and an appropriate sampling of the point spread function at the detector as well as by stellar rotation (c.f. Fig.~\ref{F:4}). In the following, instrumental limitations and the impact of the properties of the star are swiftly discussed in Subsection~2.1 and Subsection~2.2, respectively.

\subsection{Instrumental limitations} 
A crucial part of high-precision RV measurements is the wavelength calibration. This is done by adopting a stable reference spectrum, which can be a large number of dense absorption lines of a gas sealed in a cell (e.g. iodine gas cell; Butler et al. 2006), which is located in the light path of the target observation, or a dense forest of emission lines of a lamp (e.g. Th-Ar-lamp; Kerber et al.~2006). These two calibration methods have been successfully carried out in the optical. At NIR wavelengths, however, our knowledge of these calibration methods is  
still rather limited, due to the small number of comparable instrumentation in the NIR. 

A big advantage of using a gas cell is that the absorption lines of the gas cell are directly superimposed on the stellar spectrum of interest, which allows a direct monitoring of changes of the 
point spread function (PSF) of the instrument. The major disadvantage of adopting a gas 
cell is that it is hard to find a gas which has strong absorption features in a large wavelength regime, and that gases might absorb a large fraction of the weak stellar light (up to 50\%). We should 
keep in mind that for very cool dwarfs, even with very efficient NIR spectrographs mounted 
at 8m class telescopes, the integration times are still long in order to get a high SNR. 
Valdivielso et al.~(2010) present different mixtures for a gas cells providing a dense 
forest of lines in the NIR in the $H$ and $K$~band. In addition to that, Mahadevan \& Ge (2009) 
report on a mixture that provide a dense absorption spectrum in the $H$~band. Bean et al.~(2010) 
report that for their CRIRES survey on the search for planets around M-dwarfs they reach a 
RV precision of $5~{\rm m~s^{-1}}$ using a NH$_3$ gas cell in the $K$~band. 
A similar strategy for wavelength calibration is the use of telluric lines in the NIR as a natural gas cell. In the
observations, the telluric lines are superimposed 
on the stellar
spectra, and Seifahrt et al.~(2008) showed that these atmospheric lines are stable in terms of velocity
in the order of $10~{\rm m~s^{-1}}$. Figueira et al.~(2010) adopted telluric lines as a stable wavelength
reference, and showed that these lines allow to reach RV precisions even better than $10~{\rm m~s^{-1}}$ when
accounting for the changing weather conditions at the observatory during the observations.

Special lamps which produce a dense forest of emission lines have 
been used for most of the high-precision and stabilized spectrographs in the optical. For example, 
the high-precision spectrograph HARPS (Mayor et al. 2003), mounted at the 3.6~m telescope in La Silla, Chile, allows to take
 RV measurements with precisions of less than $1~{\rm m~s^{-1}}$.
The few NIR-spectrographs presently available are also equipped with lamps for wavelength calibration. 
For example, the intermediate-resolution spectrograph ($R\approx 20,000$) NIRSPEC, mounted on the 
10.2m Keck II telescope in Hawaii, USA, makes use of arc lamps of different elements,
such as Kr, Xe, Ne, and Ar, as a stable wavelength reference spectrum. In addition to that, for the high-resolution NIR spectrograph CRIRES, 
mounted on the UT1/VLT at the ESO Paranal Observatory in Chile, a Th-Ar lamp is used for wavelength 
calibration. This lamp produces more than 1800 lines in the wavelength range of $\lambda=0.9$ to 2.5~$\mu$m (Kerber et al.~2008), which is only a slightly less value than in the optical. Controversially to the gas cell, the usage of a lamp does not allow to monitor changes in the response of the instrument. 


\subsection{Limitations from the dwarfs}
Stellar rotation limits the precision that can be attained in RV measurements.  In particular, when the rotational broadening function 
is wider than the PSF, the RV precision is limited
to the stellar rotation. For this reason, it is not advisable 
to observe fast rotators with high-resolution when searching for extrasolar planets. Late M-dwarfs are the link between hotter and slowly-rotating
 M-dwarfs and Brown Dwarfs, which tend to rotate very rapidly. Reiners \& Basri (2010) found that approximately 50\% of their 63  dwarfs of spectral
 types M7-M9.5 show projected rotational velocities $v \sin i > 10~{\rm km~s^{-1}}$. This fraction of fast rotators is consistent with a study by Deshpande (2010), who investigated the properties of 36 M-dwarfs by using NIRSPEC spectra, and with Jenkins et al.~(2009), who determined the projected rotational velocities of 56 M-dwarfs.

L-dwarfs show very high rotational velocities. Del Burgo et al.~(2009) determined the rotational velocity of brown dwarfs, 
which had been measured with NIRSPEC spectrograph (McLean et al. 1998) at the Keck II telescope in Hawaii, and found that all of the investigated brown dwarfs have projected rotational velocity larger than $v \sin i \sim 16$~km~s$^{-1}$.

A well-known problem that hampers to attain high-precision RV measurements is stellar activity, in
particular, the presence of stellar spots.
These spots lead to line profile
distortions in the stellar absorption lines, which can be easily misinterpreted as a RV shift 
coming from an unseen 
companion. By observing
in the NIR, as opposed to the optical wavelengths, one observes in
a domain where the contrast between stellar spots and the stellar
disk is reduced. Consequently, the semi-amplitude of a RV signal produced by a stellar spot is
large at optical wavelengths, and becomes smaller towards infrared wavelengths (e.g. Mart\'in et al.~2006).
Reiners et al.~(2010) and Barnes et al.~(2011) carried out extensive simulations in order to investigate 
the effect of a stellar spot on low mass dwarfs on RV measurements at NIR wavelengths. 
They find that even for low temperature differences between stellar surface and spot, 
they produce spurious RV-shift with semi-amplitudes larger than 3~m~s$^{-1}$. For larger 
temperature differences, stellar spots can produce spurious shifts with semi-amplitudes larger 
than 100~m~s$^{-1}$. We note that  an effective method to distinguish between spots or real
 RV-shifts induced by orbiting planets is the analysis of the line bisectors (e.g. Hatzes et al.~1997).

\section{Simulations}

\subsection{Theoretical models for M- and L-dwarfs}

M- and L-dwarfs are very faint in the optical but brighter in the
infrared as their spectral energy distribution peaks at around 1.2 - 1.3$~\mu {\rm m}$. In the NIR,
the spectra of late
M-dwarfs show strong alkali lines (K I, Na I, Mn I, Al I), metal hydride bands (FeH, CaH) (Mart\'in et al.,
1999), including water, CO, FeH and VO bands (Mclean et al. 2007, Cushing et al. 2005).
In the last few years there has been an emergence of infrared spectroscopic studies which have shed
light on their physical characteristics (temperature, mass, radii) - e.g. Deshpande (2010), Jackson et al. (2009), and Jenkins et al.~(2009).

Early L-dwarfs show a mixture of atomic and molecular bands, the most prominent being
the neutral alkali lines (Na I, K I, Rb I, Cs I, and sometimes Li I), oxide bands TiO
and VO, hydride bands CrH and FeH, and CaOH. Going to cooler L -dwarfs, the ground-state Na I
and K I lines grow tremendously in strength; the molecules MgH, CaH, CrH,
and FeH get stronger, whereas the oxides TiO and VO largely
disappear. By late-L and early-T, H$_2$O increases in strength, the neutral
alkali lines are still strong, and the hydrides are much reduced in prominence (Kirkpatrick et al.~1999).

Del Burgo~et al.~(2009, 2010) determined the physical properties (in particular, effective temperature, surface gravity and rotational broadening) 
of M- and T-dwarfs from the comparison of stellar atmosphere synthetic models and J-band high-resolution 
(resolving power of $\sim$ 20,000) spectra obtained with the NIRSPEC spectrograph at the Keck II telescope in Hawaii. 
For their analysis of M-dwarfs with $T_{\rm eff}\ge3000$~K, they used the general-purpose stellar atmosphere PHOENIX code 
(Hauschildt \& Baron, 1999; Baron et al. 2003)
In particular, the version 16 that includes a number of improvements compared to 
previous versions, such as a complete new equation of state for ions, molecules and condensation, updated opacity databases, and improved line profiles for atomic lines. 
For $T_{\rm eff}$ below 3000~K, del Burgo~et al.~(2009b) used
the Drift-PHOENIX code to produce theoretical spectra of late M- and T-dwarfs, which satisfactorily reproduced the observed spectra.
That code is a merger of the PHOENIX code and the dust model Drift 
(Helling et al.~2008a; Witte et al.~2009). The dust grains are composites and yield improved opacities in contrast to the grains in 
earlier models, and the use of a kinematic, phase-non-equilibrium dust formation model avoids an overestimated condensation/evaporation 
(Helling et al. 2008b).

 Fig.~1 shows the theoretical Drift-PHOENIX spectrum of an M9.5-dwarf as well as for an L-dwarf with an effective temperature of  $T_{\rm eff}=1800$~K in the four atmospheric windows ($Y$, $J$, $H$ and $K$ bands).

\begin{figure}[t!]
   \centering
    \includegraphics[width=9cm]{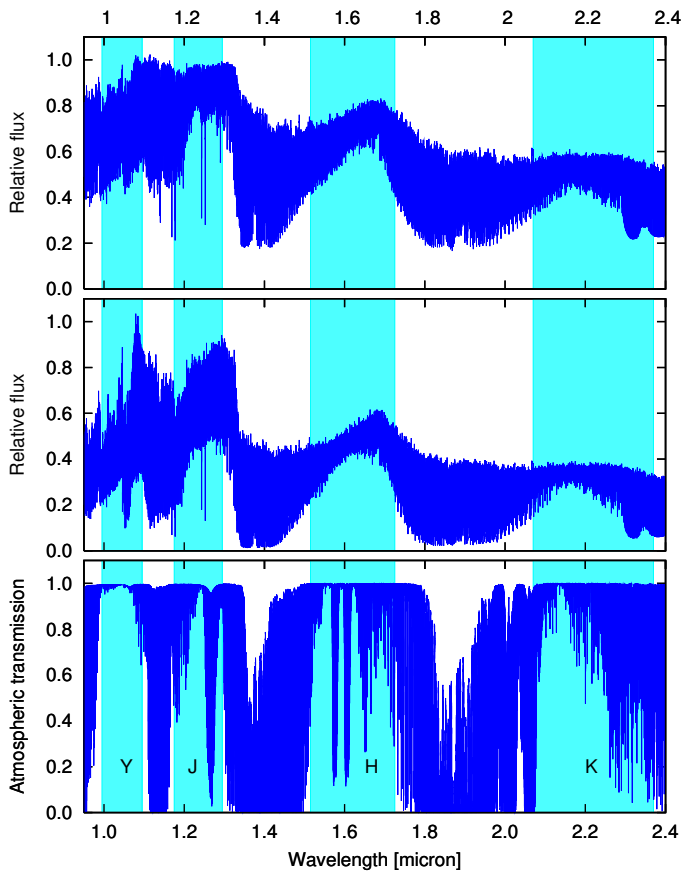}
\caption{Bottom: Theoretical transmission spectrum of the Earth's atmosphere, calculated for the Observatorio de los Muchachos, La
	  Palma, Spain. 
	  The four atmospheric windows
	  ($Y$, $J$, $H$, $K$~band) are indicated. 
	  These four transmission bands are not entirely telluric-free:
	  While the $Y$~band shows only little contamination, the $J$, $H$, and $K$~bands show significant telluric contamination. ~~~~~~
	  Middle: Theoretical spectrum of a hot L-dwarf in the wavelength region between 0.9 and 2.4$~\mu {\rm
	  m}$, which was generated with the Drift-PHOENIX code for the following parameters: $T_{\rm eff}=1800$~K, 
	  $\log~g=4.5~{\rm cm~s^{-2}}$ and  solar metalicity. 
The positions of the four atmospheric windows
	  ($Y$, $J$, $H$, $K$~band) are indicated. ~~~~~~%
Top: Theoretical spectrum of a very cool M-dwarf in the wavelength region between 0.9 and 2.4$~\mu {\rm
	  m}$, which was generated with the Drift-PHOENIX code for the following parameters: $T_{\rm eff}=2200$~K, 
	  $\log~g=4.5~{\rm cm~s^{-2}}$, and  solar metalicity. 
The positions of the four atmospheric windows
	  ($Y$, $J$, $H$, $K$~band) are indicated. While the model spectrum in the $Y$~band shows a forest of deep absorption lines, the spectrum in the $J$ is characterized by shallow features, apart from the strong potassium doublet around 1.25~$\mu$m. Close to the blue and red edge of the $H$~band, deep absorption lines are observed, while the center of this band shows shallow absorption lines. In the $K$~band, the most promising wavelength region is red-wards of 2.29~$\mu$m, where a forest of deep CO-absorption lines is present.	  }
          \label{F:1}
   \end{figure}

\subsection{Creating the data sets}
To investigate in which atmospheric window we can achieve the highest RV
precision for the late M and L-dwarf, we create artificial data sets which closely resemble real
observations. The free parameters in these data sets were: wavelength band ($Y$, $J$, $H$, or $K$ band), the stellar rotation $v \sin i$, the spectral resolving power, and the signal-to-noise ratio (SNR) per spectrum.

For the simulations, we adopt two high-resolution Drift-PHOENIX models with the following parameters: For the M-dwarf with $T_{\rm eff}=2200$~K and $\log~g=4.5~{\rm cm~s^{-2}}$, while for the L-dwarf with  $T_{\rm eff}=1800$~K and $\log~g=4.5~{\rm cm~s^{-2}}$ (cf. Fig.~1). The flux of the stellar model is in energy units.  In order to convert the flux of the model 
into photon units as recorded by the detector, we multiply the flux with the its corresponding wavelength value $\lambda$ throughout the whole spectral range of the model. In the next step, we Doppler shift this template spectrum by a specified velocity. We further account for stellar rotation by convolving the stellar spectrum with a rotational profile, which we computed using the approach described in Gray~(2005) for a chosen rotational broadening velocity $v \sin i$ and a limb darkening coefficient of $\epsilon= 0.6$. 

As the stellar light passes through the Earth's atmosphere, it is partially absorbed by molecules in the atmosphere, 
such as water and methane. Mathematically, the stellar spectrum is multiplied with the transmission spectrum of the Earth's atmosphere.
The transmission spectrum was calculated by adopting the Line-By-Line
Radiative Transfer Model (LBLRTM) code, which is based on the FASCODE
algorithm (Clough et al.~1992). As molecular data base we adopt HITRAN (Rothman et al.~2005). 
We note that we carried out a comparison study between the theoretical transmission spectra and 
observed mid-resolution ($R\sim20,000$)spectra, taken with NIRSPEC (McLean et al.~1998) at Keck II in Hawaii, USA, 
and found that the theoretical spectra reproduce the observations in an excellent way. These results will be presented in 
a forthcoming paper (Rodler et al., in prep.).

For this paper, we calculate the telluric contamination for the Observatorio de los Muchachos in La Palma, Spain, which is located at 2400~m above sea level.
The strengths of the telluric lines are sensitive to the local elevation in the sky. In our simulations, we assume that we observe our star at low
air-masses at an elevation of 65 degrees above horizon. In Fig.~\ref{F:1} (bottom panel) we show the telluric transmission spectrum and mark the four atmospheric windows of interest. This figure depicts  that none of these four atmospheric windows is completely telluric-line-free. For this reason, in each band telluric lines appear superimposed on the stellar spectrum. It is obvious that when finally calculating the RVs, these lines needs to be removed from the stellar spectra.

Before the stellar light is recorded at the detector, the light is collected in the telescope and processed in the spectrograph. In the instrument, the stellar spectrum (and the telluric spectrum superimposed on it) is degraded in terms of spectral resolving power and binning on the detector. Mathematically, the stellar spectrum is convolved with the point-spread function (PSF) of the spectrograph. Usually, the PSF of the spectrograph is close to a Gaussian function, but it can be slightly asymmetric (e.g. Endl et al.~2000, Bean et al.~2010).
In our simulations, we adopted a Gaussian function as the PSF of the spectrograph. We created 
data sets for different spectral resolving powers, namely $R=\lambda/\Delta\lambda=20,000, 40,000, 60,000$ and $80,000$, and assumed that the instrument
collects the same number of photons at each instrumental setting.
The resulting spectra were then interpolated onto a
 chosen detector grid, with a 2 pixel sampling per resolution element, 
 were the spectral resolving power was kept constant
 across the full spectral range. 
 
In the final step, we add Poisson noise to the data according to the certain SNR in the stellar continuum at a selected wavelength $\lambda$.
We assume that the detector has the same
sensitivity throughout the observed wavelength region, and we factor the spectral flux distribution of the dwarf in. To this end, we 
simply scale the spectrum so that the 
spectral flux distribution information is conserved and the maximum flux is the squared value of the desired SNR. 
The actual SNR per spectral pixel is then determined as the square root of the flux at that pixel. 
When finally adding Photon noise to the data, this approach makes sure that regions of telluric lines as well as stellar 
absorption lines have a lower SNR, respectively.

\subsection{Analysis of the data sets}
 
Once having created our ``observations'', we apply the same analysis we would follow when dealing with truly observed spectra. For the treatment of the telluric contamination in our data analysis, we investigate two scenarios: (a) A complete and perfect removal of the telluric lines in order to see the highest RV precision possible. The only 
indicator for the removed telluric lines is a 
reduced SNR in those regions.  
In the more realistic approach (b), we chose a similar approach as presented in Reiners et al.~(2009) and mask out those telluric lines with absorption depths $\ge3$\% at a
spectral resolving power of $R=80,000$. Additionally, we also discard the close vicinity ($\pm30$~km~s$^{-1}$) to these lines from further analysis in
order to account for the velocity shifts produced by the barycentric motion of the Earth.

In order to avoid effects of the dispersion (e.g. the spectral lines get broader towards larger wavelengths), 
we convert the dispersion axis from wavelength $\lambda$ to velocity $v$. To this end, we apply the following Equation~\ref{E:1}:
\begin{equation}       \label{E:1}
v =  \frac{{\rm c}}{\lambda}{\rm d}\lambda = {\rm c} \ln \lambda ,
\end{equation}
where c is the speed of light in a given unit, which is then the unit of the velocity grid.

In the next step, we subdivide the spectrum into equidistant pixel chunks of size 200 pixel each. We give each chunk a weight $w_i$ according to the 
average SNR of the ``observation''.
For each wavelength chunk, we 
cross-correlate the ``observation'' with a {\it template}, which is nothing but a high-SNR reference spectrum of the ``observation''.
Then, we calculate the relative RV with respect to the {\it template}. Those chunks that show RV values which are clearly outliers are 
discarded from further analysis. Finally, we determine the error of the RV-measurement as:
\begin{equation}       \label{E:2}
\sigma_{\rm RV} = \frac {{\rm rms}~w_i}{\sqrt{ \sum w_i}},
\end{equation}
where $w_i$ is the weight of the unrejected chunk $i$.


\section{Results}

\begin{figure}[t!]
   \centering
   \includegraphics[angle=-90,width=9cm]{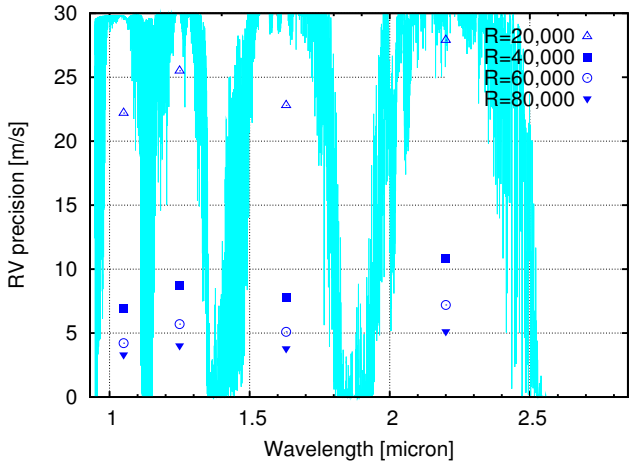}
      \caption{Expected RV precisions of an M9.5-dwarf  with a temperature of 2200~K and a projected rotational velocity of $v \sin i = 5$~km~s$^{-1}$ in the different NIR wavelength
	  bands ($Y$, $J$,
 $H$ and $K$) for different spectral resolving powers. For these simulations, we factored in the spectral flux
 distribution of the M9.5-dwarf, and assumed a homogeneously sensitive detector throughout the NIR.
 The gray spectrum in the background represents the transmission of the Earth's atmosphere (the value 30
 in the y-axis corresponds to 100\% transmission). The SNR in the brightest regions in the stellar continuum 
 of the M-dwarf was set to be 90 for a
 spectral resolving power of $R=80,000$. The SNR values and the RV precision values for the different
 wavelength bands and spectral resolving powers are given in Table~1. Although the spectral flux distribution of such a cool
 dwarf peaks around 1.2~$\mu$m, the highest radial velocity precision is attained in the $Y$~band.}
          \label{F:2}
   \end{figure}
   
\begin{figure}[t!]
   \centering
   \includegraphics[angle=-90,width=9cm]{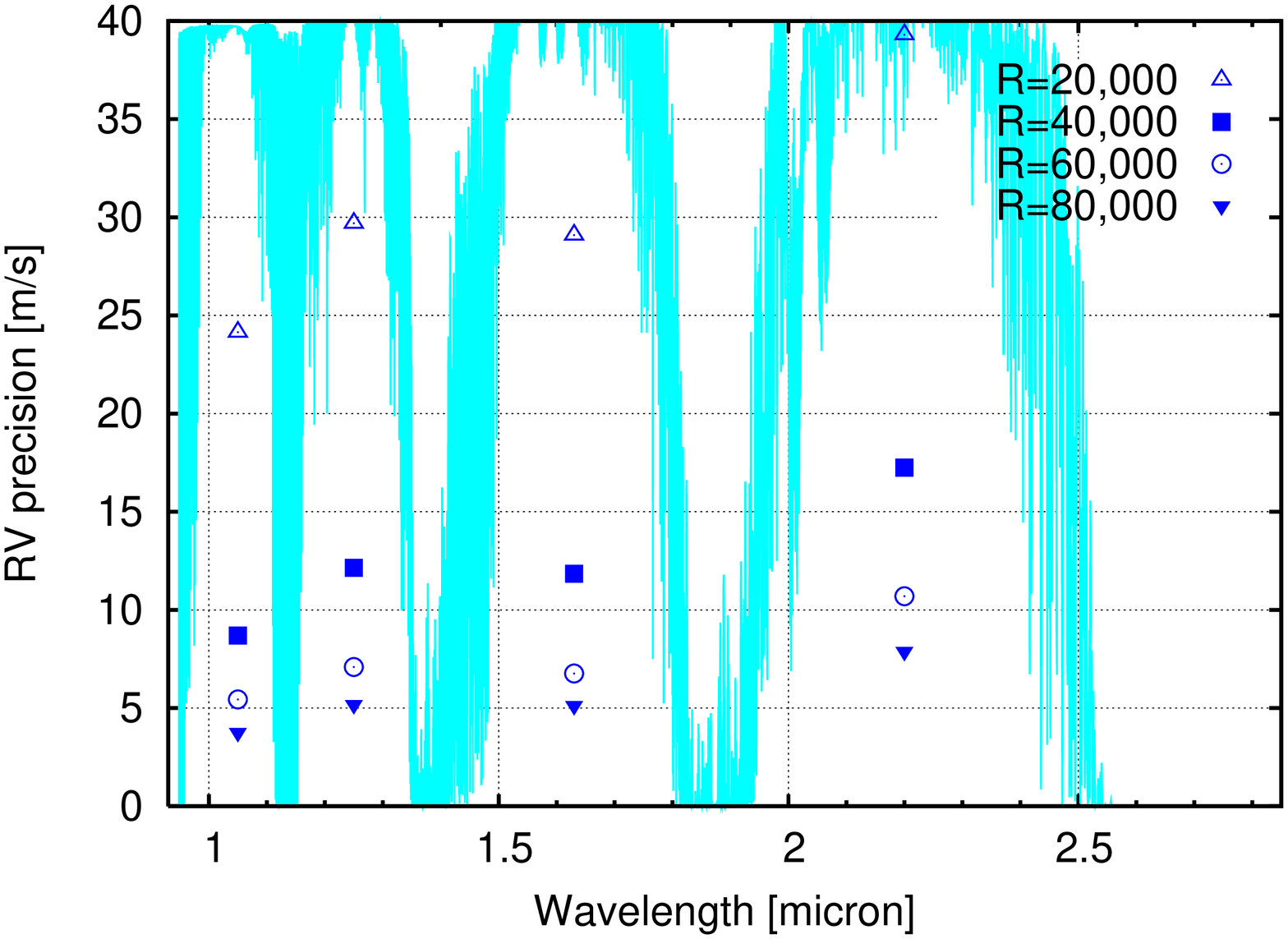}
      \caption{Same as Fig.~2, but for scenario B, where we masked out regions of
  heavy telluric contamination (i.e. 3\% or more of telluric absorption at R=80,000) in our analysis. The gray spectrum in the background represents the transmission of the Earth's atmosphere (the value 40
 in the y-axis corresponds to 100\% transmission). The highest radial velocity precision is attained in the $Y$~band.}
          \label{F:22}
   \end{figure}

\begin{figure}[t!]
   \centering
   \includegraphics[angle=-90,width=9cm]{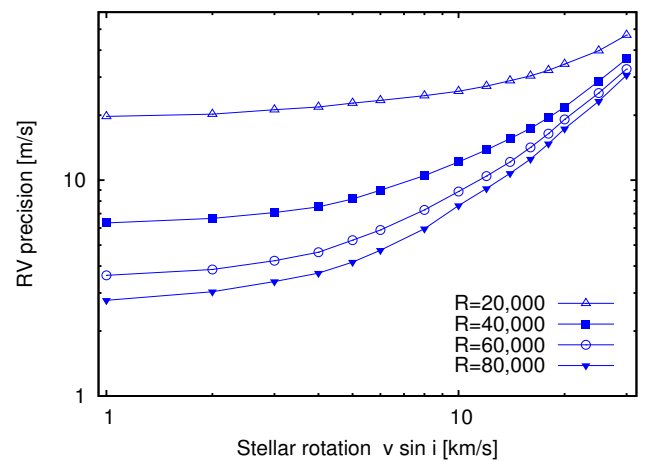}
      \caption{Achievable radial velocity of the very cool M-dwarf as a function of the stellar rotation $v \sin i$ and the spectral
	  resolving powers. For the simulations, we created data sets for the $Y$~band with the average SNR
	  given in Table~1.
	  in the stellar continuum. The free parameters for the simulation were the stellar rotation and the
	  spectral resolving power.}
          \label{F:4}
   \end{figure}

\subsection{M9.5-dwarf}
We analyzed data sets of the M9.5-dwarf in the four different atmospheric windows and four different 
spectral resolving powers, which had the following configuration: a 
stellar rotation of $v \sin i = 5~{\rm km~s^{-1}}$, and a
${\rm SNR}=90$ in stellar continuum at the peak of the spectral energy distribution 
of the dwarf at a spectral resolving power of $R=80,000$ in order to account for the intrinsic faintness of these class of objects. 
The corresponding SNR ratios in the other wavelength bands and spectral resolving powers are given in Table~1.

Figure~\ref{F:2} shows the analysis for the scenario that the telluric lines were completely and perfectly removed (Scenario A), while Figure~\ref{F:22} depicts the analysis with
telluric regions discarded from analysis (Scenario B). The highest RV precision is 
attained in the $Y$~bandwidth with $v=3.4$~m~s$^{-1}$ and $v=3.9$~m~s$^{-1}$ for a spectral resolving power of $R=80,000$, respectively for the
Scenarios A and B. We note that in our simulations we show the best case, i.e.
perfect wavelength solution and a complete wavelength coverage throughout the transmission bands without inter-order gaps.

Fig.~\ref{F:4} shows the results of our studies of how the RV precision  depends on the stellar 
rotation $v \sin i$ and the spectral resolving power. We carried out the simulations in the $Y$~band with the SNR values 
in the stellar continua given in Table~1. In this figure we see that for slow rotating stars ($v \sin i < 10~{\rm km~s^{-1}}$),
observing at higher resolution allows to take RV measurements with a lower RV error. However, for fast
rotators, there is hardly no improvement in terms of achievable RV precision between the spectral resolving powers
of $R=60,000$ and $80,000$, since the width of the rotational broadening function is roughly the same than the  width of
the PSF of the spectrograph.

We furthermore investigated what kind of planets could be discovered with the attained RV precisions. We have to remind the reader that in our simulations, we did not account for wavelength calibration errors nor other instrumental issues, which could introduce noise to the data and lower the RV precisions. Consequently, the following numbers denote the best case scenario and the RV precision might be lower in the real case.

By employing the following equation, we determined the minimum mass of the planet as
\begin{equation} \label{xoxoequ:5}
m_{\rm p,min} = K_{\star} m_{\star}^{1/3} \left ( \frac{P}{2 \pi {\rm G}}\right ) ^{2/3},  
\end{equation}
where $K_\star$ the RV semi-amplitude of the reflex motion of the star, $m_\star$ is the stellar mass, $P$ the orbital period of the planet, and G the gravitational constant. We considered only the case of a circular orbit. We set the mass of the very cool M-dwarf to $m_\star=0.075~{\rm M}_\odot$, and calculated the minimum masses of the planet with respect to the orbital period $P$ and $K_\star$.   
Fig.~\ref{F:5} shows the results we determined for different spectral resolving powers in the optimum wavelength band ($Y$ band). For example, if we obtain a spectrum of an M9.5-dwarf with a rotational broadening of $v \sin i= 5$~km~s$^{-1}$, in the $Y$ band with a SNR of 80 in the stellar continuum at a spectral resolving power of $R\ge 60,000$, we are able to detect short period planets with minimum masses $m_{\rm p,min} \sim 2~{\rm M_\oplus}$. 

\begin{table}
\begin{center}
\caption{Radial velocity precisions of the dwarfs that can be attained for spectral resolving powers $R$ and SNR in the stellar continua.
Scenario~A shows the data analysis case of a complete and perfect removal telluric contamination, while
Scenario~B lists the RV precisions for the data analysis,
where the telluric lines of a absorption depth of $\ge3\%$ were masked out.
} \label{xoxo:T4}
\begin{tabular}{ccccccccc}
\hline
\hline
$R$ &\multicolumn{4}{c}{SNR} & \multicolumn{4}{c} {RV precision (m s$^{-1}$)}\\
& $Y$ & $J$ & $H$ & $K$ & $Y$ & $J$ & $H$ & $K$ \\
\hline
&\multicolumn{8}{c}{M9.5V-dwarf ($T_{\rm eff}=2200$~K) + Scenario~A} \\
\hline
20,000& 139 & 180 & 171 & 152 & 22.2 & 25.5 & 22.8 & 27.9\\
40,000&  98 & 127 & 121 & 108 & 6.9 & 8.7 & 7.8 & 10.8 \\
60,000&  80 & 104 & 99 & 88 & 4.2 & 5.7 & 5.1 & 3.8\\
80,000&  70 & 90 &  85 & 76 & 3.3 & 4.0 & 3.8 & 5.1 \\
\hline
&\multicolumn{8}{c}{M9.5V-dwarf ($T_{\rm eff}=2200$~K) + Scenario~B} \\
\hline
20,000& 139 & 180 & 171 & 152 & 24.2 & 29.7 & 29.1 & 39.3\\
40,000&  98 & 127 & 121 & 108 & 8.7 & 12.2 & 11.9 & 17.3 \\
60,000&  80 & 104 & 99 & 88 & 5.4 & 7.1 & 6.8 & 10.7 \\
80,000&  70 & 90 &  85 & 76 & 3.8 & 5.2 & 5.1 & 7.9\\
\hline
&\multicolumn{8}{c}{L-dwarf ($T_{\rm eff}=1800$~K) + Scenario~A} \\
\hline
20,000& 68 & 90 & 88 & 84  & 83.2 &  73.8 & 92.5  &  92.5 \\ 
40,000&  47 & 64 & 63 & 59 & 60.6 &  56.8 & 66.9  &  66.9 \\ 
60,000&  38 & 52 & 52 & 50 & 49.0 &  45.1 & 53.5  &  53.5 \\ 
80,000&  33 & 45 & 44 & 42  &42.3 &  37.5 & 57.0  &  57.0 \\ 
\hline
&\multicolumn{8}{c}{L-dwarf ($T_{\rm eff}=1800$~K) + Scenario~B} \\
\hline
20,000& 68 & 90 & 88 & 84  &  87.6 & 77.0 & 109.2& 158.3\\
40,000&  47 & 64 & 63 & 59 &  65.9 & 61.8 & 84.2 & 114.5\\
60,000&  38 & 52 & 52 & 50 &  55.3 & 50.9 & 70.0 & 95.6 \\
80,000&  33 & 45 & 44 & 42  & 47.3 & 41.4 & 60.7 & 79.8 \\
\hline

\end{tabular}
\end{center}
\end{table}

\begin{figure}[t!]
   \centering
   \includegraphics[angle=-90,width=9cm]{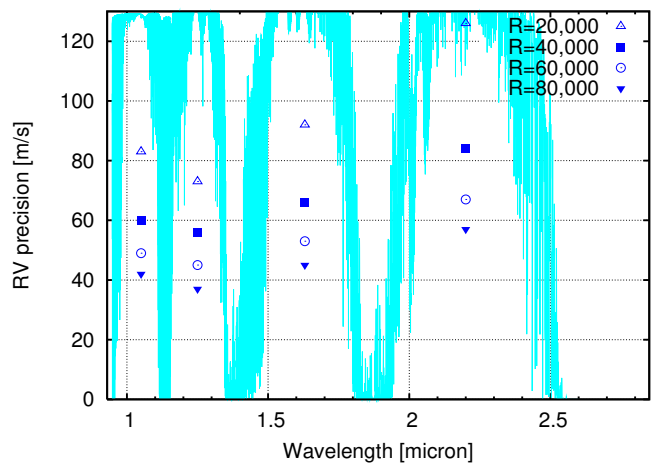}
      \caption{Expected RV precisions of hot brown dwarf with a temperature of 1800~K and a projected rotational velocity of $v \sin i = 20$km~s$^{-1}$ in the different NIR wavelength
	  bands ($Y$, $J$,
 $H$ and $K$) for different spectral resolving powers. For these simulations, we factored in the spectral flux
 distribution of the brown dwarf, and assumed a homogeneously sensitive detector throughout the NIR. 
 The SNR in the brightest regions in the continuum of the brown dwarf was set to be 45 for a
 spectral resolving power of $R=80,000$. The SNR values and the RV precision values for the different
 wavelength bands and spectral resolving powers are given in Table~1.
 The gray spectrum in the background represents the transmission of the Earth's atmosphere (the value 130
 in the y-axis corresponds to 100\% transmission). The highest radial velocity precision is attained in the $J$~band.}
          \label{F:3}
   \end{figure}

\begin{figure}[t!]
   \centering
   \includegraphics[angle=-90,width=9cm]{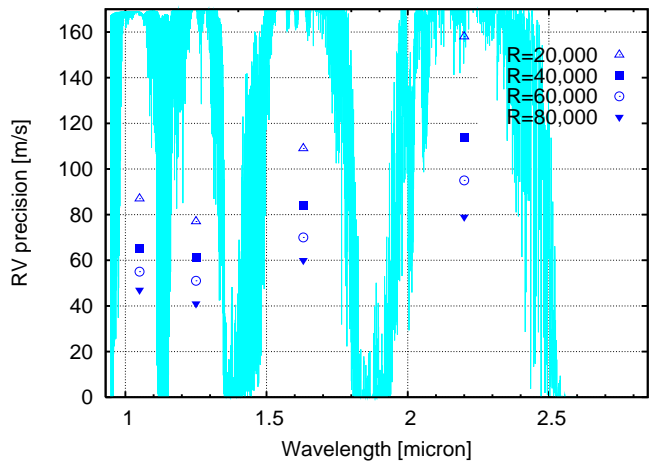}
      \caption{Same as Fig.~5, but for scenario B, where we masked out regions of
  heavy telluric contamination (i.e. 3\% or more of telluric absorption at R=80,000) in our analysis.  
   The gray spectrum in the background represents the transmission of the Earth's atmosphere (the value 170
 in the y-axis corresponds to 100\% transmission). The highest radial velocity precision is attained in the $J$~band.}
          \label{F:33}
   \end{figure}

\subsection{L-dwarf}
We analyzed data sets of the L-dwarf with a temperature of 1800~K in the four different atmospheric windows and four different spectral resolving powers, which had the following configuration: In order to account for the rapid rotation among brown dwarfs, we set the
stellar rotation to $v \sin i = 20~{\rm km~s^{-1}}$. In addition to that, to account for the relative faintness of hot L-dwarfs by adopting a
${\rm SNR}=45$ in the stellar continuum at the peak of the spectral energy distribution of the L-dwarf at a
 spectral resolving power of $R=80,000$. The corresponding SNR ratios in the other wavelength bands and spectral resolving powers are given in Table~1.

Fig.~\ref{F:3} shows the results of Scenario~A where the telluric lines were completely 
and perfectly removed from the spectra, while in Fig.~\ref{F:33} we show the results of Scenario~B where the regions of telluric contamination were masked out and discarded from further analysis.
As result, for both scenarios we find that the highest RV precision for the L-dwarf  is achieved in
the $J$. The RV precisions for the different wavelength regions and spectral resolving powers are given in
Table~1.

\begin{figure}[t!]
   \centering
   \includegraphics[angle=-90,width=9cm]{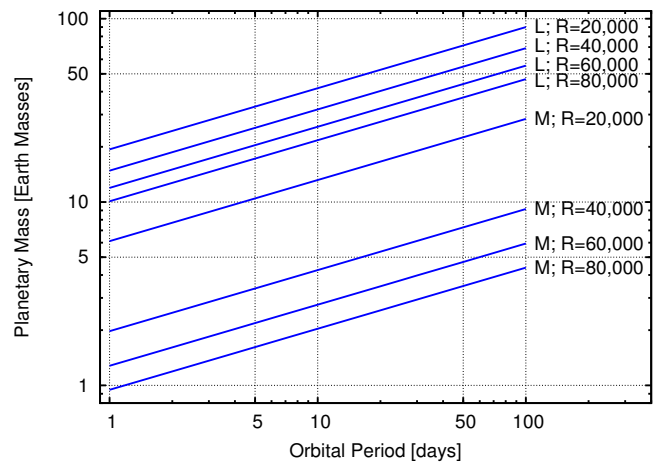}
      \caption{Minimum masses of the planetary companions with respect to the orbital periods and the RV precisions, 
      which are a function of the spectral resolving power $R$. For the late M-dwarf (in the plot marked with ``M'') we
       assumed a mass of $0.075$~M$_\odot$ and a projected rotational velocity of $v \sin i= 5~{\rm km~s^{-1}}$. We calculated 
       the masses by adopting Equation~\ref{xoxoequ:5} and the RV precision we had determined for different spectral resolving 
       powers (R=20,000, 40,000, 60,000 and 80,000) in the $Y$ band with a SNR of 80 in the spectral continuum. We demonstrate that planetary companions with short periods and slightly higher masses than the Earth can be detected when taking observations at high spectral resolving powers.
For the L-dwarf (marked with ``L'') with a temperature of 1800~K, we assumed a mass of $0.07$~M$_\odot$ and a projected rotational velocity of $v \sin i= 20~{\rm km~s^{-1}}$. As result, we find that hot Neptunes can be detected when observations are taken in the $J$ band, with a spectral resolving power of $R\ge60,000$ and a SNR~ of 45 in the spectral continuum.  }
          \label{F:5}
   \end{figure}

Adopting Equation~\ref{xoxoequ:5} and the RV precisions we had previously determined for the different spectral resolving powers in the $Y$ band, we calculated the minimum masses of possible planetary companions to the L-dwarf with respect to the orbital period. In 
Fig.~\ref{F:5} it is shown that hot Neptunes could be detected with minimum masses of $m_{\rm p,min} > 20~{\rm M_\oplus}$ and periods $\le 5$ days, when observations are taken at a spectral resolving power of $R\ge 60,000$, and a conservative value for the mass of the L-dwarf with $m_\star=0.07$~M$_\odot$ is assumed.

\section{Summary and discussion}

We have investigated which atmospheric window in the NIR (i.e. $Y$, $J$, $H$, and $K$~band) is best-suited for the search for rocky planets around
a very cool M-dwarf and L-dwarf with a temperature of $T_{\rm eff}=2200$~K and 1800~K, respectively. 
We carried out the simulation for two scenarios in order to investigate the issue of the telluric
contamination: Scenario~A denotes
the analysis with a complete and perfect removal of telluric lines from the spectra in order to see what RV
precision would be possible in the best case for different instrumental settings, while in the more realistic Scenario~B 
regions at and around telluric absorption lines of $\ge3\%$ depth are masked out and discarded from the RV
determination process.

We note that the RV precisions presented in this paper are estimations, since our simulations are based on assumptions:
We are using theoretical PHOENIX-Drift model spectra to describe the dwarfs. These synthetic models may lack of some faint absorption features or underestimate their abundances. 
Conversely, some instrumental/calibration aspects, such as the errors coming from the wavelength calibration or noise coming from the detector (e.g.
read-out noise, sky background variations, and thermal noise) as well as possible stellar activity aspects are not 
taken into account in our estimations
which would rise the uncertainties and lower the RV precision. 

For the M9.5 dwarf, we find for both scenarios for the treatment of the telluric contamination that the 
highest RV precision can be achieved in the $Y$~band. Our results support the conclusions drawn by Reiners et al.~(2010) for their choice of hotter M-dwarfs ($T_{\rm eff}=3500$~K, 2800~K and 2600~K). We demonstrate that planetary companions with short periods and slightly higher masses than the Earth can already be detected when taking observations in the $Y$~band at spectral resolving powers of $R\ge60,000$. Furthermore, we have studied the dependence of the stellar rotation 
and the spectral resolving power on the achievable RV precision. 
For all the rotational velocities the highest spectral resolving power of 80,000 yields the
 smallest errors in the RV measurements. However, for fast rotators ($v \sin i \ge 10~{\rm km~s^{-1}}$), we see hardly any difference in RV precision between 
 a spectral resolving power of 60,000 and 80,000. 

For the L-dwarf with a temperature of 1800~K, the highest RV precision in both scenarios is attained 
in the $J$ band. Since these objects are intrinsically very faint and tend to rotate rapidly, 
the achievable RV precisions are about a magnitude higher than those for late M-dwarfs. 
Nevertheless, we demonstrate that at least hot Neptunes could be discovered around L-dwarfs 
when taking observations at spectral resolving powers of $R \ge 60,000$.
 
 We furthermore compared the RV precisions between both scenarios of telluric lines treatment:
In the $Y$ band,
the difference in RV precision between Scenario~A and B  is relatively small, due to the low contamination in
this wavelength band. In the $J$ and $H$~band, we encounter relatively moderate telluric contamination, which
affect the RV determination process. However, in the $K$ band, heavy telluric contamination throughout the
whole wavelength band severely affects the RV precision (c.f. Table~1 and Fig.~\ref{F:1}).

 Our studies bring out the need for high-resolution cross-dispersed spectrographs to search for rocky planets and/or to follow up transit candidates around cool M-dwarfs and L-dwarfs. 
 In order to avoid problems with the wavelength calibration and to ensure the stability required, we suggest that such spectrographs should 
 only have one observing mode as well as allow to take Th-Ar exposures simultaneously with the target observations. 
 The spectrograph should simultaneously cover the $Y$, $J$ and $H$~band, but avoid the 
 $K$~band. Extending the spectrograph to the latter atmospheric window would require severe 
 cooling and different optical parts (e.g. different fibers which have a high optical throughput in the $K$~band). 
 Since roughly the half of the cool M-dwarfs as well as most of the L-dwarfs are fast rotators, a spectral resolving power of $R\sim60,000$ is sufficient to achieve
 high RV precisions.
  A practical advantage of a lower resolution with respect to the spectral resolving power of $R=80,000$ would be that inter-order gaps between the spectral orders would be 
  kept relatively small, and consequently the wavelength coverage would be higher. 

The results of our simulations led to a re-design of the high-resolution part of the NIRINTS spectrograph (formerly NAHUAL; Mart\'in et
al.~2005), to be mounted on the GTC at the Observatorio de los Muchachos in La Palma, Spain. This spectrograph is planned to have a spectral resolving power of  $R\sim61,000$ and to cover the entire $Y$, $J$ and $H$~bands. 
 Main science case for this spectrograph is 
to search for and to follow up rocky planets around cool M-dwarfs and L-dwarfs.

Recently, further projects to build high-resolution spectrographs covering one or more atmospheric windows in the NIR  have been launched. For example, the project CARMENES (Quirrenbach
et al.~2009), 
with a proposed spectral resolving power of $R\sim85,000$, covering the $Y$ and $J$~bands, to be mounted at the 3.5~m telescope of the Calar Alto Observatory in Spain. Science case for this spectrograph is to monitor (massive) M-dwarfs. The Habitable Zone Planet Finder (HZPF) is a proposed instrument for the 10m class Hobby Eberly telescope at the McDonald observatory in Texas, USA, which will provide a spectral resolving power of $R\sim50,000$ and will cover the $Y$, $J$ and parts of the $H$~bands (Mahadevan et al.~2010). This instrument will be capable of discovering low mass planets around M dwarfs.

  \begin{acknowledgements}
This work has been supported by the Spanish Ministerio de Eduaci\'on  
y Ciencia through grant AYA2007-67458. FR is very grateful to  Andreas Seifahrt, who helped to install LBLRTM and HITRAN to compute transmission spectra of the Earth's atmosphere.
Furthermore, we want to thank the anonymous referee for very important suggestions.

   \end{acknowledgements}


\end{document}